\documentclass[preprint,12pt]{elsarticle}

\usepackage{float}
\usepackage{pdflscape} 
\usepackage{adjustbox}
\usepackage{tikz}
\usepackage{array}
\usepackage{longtable,tabu}
\usepackage{afterpage}
\usepackage{amssymb, amsfonts}
\usepackage{graphicx}
\usepackage{chngcntr}
\usepackage{amsmath}
\usepackage{booktabs,tabularx, colortbl}
\usepackage{xcolor} 
\usepackage{multirow}
\usepackage{algorithm2e}
\usepackage[T1]{fontenc}
\usepackage{ragged2e}
\usepackage{natbib}[numbers,sort&compress]
\usepackage{url}
\usepackage{hyperref}

\SetKwComment{Comment}{/*}{*/}

\usepackage{amssymb}
\usepackage{amsmath}

\journal{IJHCS}

\makeatletter
\def\ps@pprintTitle{%
 \let\@oddhead\@empty
 \let\@evenhead\@empty
 \let\@oddfoot\@empty
 \let\@evenfoot\@empty
}
\makeatother

\begin{document}

\begin{frontmatter}



\title{Navigating with Haptic Gloves: Investigating Strategies for Horizontal and Vertical Movement Guidance}

\author[vt]{Mahdis Tajdari}
\author[jmu]{Jason Forsyth}
\author[vt]{Sol Lim}
\affiliation[vt]{organization={Department of Industrial and Systems Engineering, Virginia Tech},
            addressline={1185 Perry Street},
            city={Blacksburg},
            state={VA},
            postcode={24061},
            country={USA}}
\affiliation[jmu]{organization={Department of Engineering, James Madison University},
            addressline={800 South Main Street},
            city={Harrisonburg},
            state={VA},
            postcode={22807},
            country={USA}}



\begin{abstract}
    \small	
    Navigating peripersonal space requires reaching targets in both horizontal (e.g., desk) and vertical (e.g., shelves) layouts with high precision. 
    We developed a haptic glove to aid peripersonal target navigation and investigated the effectiveness of different feedback delivery methods.
    Twenty-two participants completed target navigation tasks under various conditions, including \textit{scene layout} (horizontal or vertical), \textit{guidance approach} (two-tactor or worst-axis first), \textit{guidance metaphor} (push or pull), and \textit{intensity mode} (linear or zone) for conveying distance cues.
    Task completion time, hand trajectory distance, and percentage of hand trajectory in critical area were measured as the performance outcome, along with subjective feedback.
    Participants achieved significantly faster task completion times and covered less hand trajectory distance in the horizontal layout, worst-axis first approach, and pull metaphor conditions. Additionally, male participants demonstrated superior performance and reported lower levels of frustration compared to their female counterparts throughout the study. Intensity mode had no significant effect on the results.
    In summary, vibrating one tactor at a time (worst-axis first) and using the pull metaphor were the most effective way of delivering vibrotactile feedback for peripersonal target navigation in both horizontal and vertical settings.
    Findings from this work can guide future developments of haptic gloves for people with vision impairments, environments with vision limitations, and for accessibility and rehabilitation applications.

\end{abstract}





\end{frontmatter}



\section{Introduction}
\label{introduction}

   
Exploring items within arm's reach, known as peripersonal navigation~\cite{brozzoli2011peripersonal}, is a common daily task. 
However, this task can be challenging in certain circumstances (e.g., crowded settings such as grocery stores or cluttered book shelves, or in situations with poor visibility) or among visually impaired individuals. 
To aid in peripersonal search and navigation, different types of sensory display technologies have been developed, such as by using visual feedback through augmented reality (AR) technology~\cite{zhao2016cuesee-ar-visual, biocca2006attention-ar-visual, warden2022visual-ar}, auditory~\cite{geronazzo2016-auditory, parseihian2012soun-auditory}, and haptic feedback~\cite{wei2022object-haptic, lehtinen2012dynamic-haptic}. 
Among these, haptic feedback has been mostly explored for peripersonal navigation, as it can provide a personalized experience that is not dependent on vision conditions and is not interfered by environmental noise, contrary to visual or auditory feedback modes~\cite{castle-2004tactile}. 

Many prior studies on using vibrotactile feedback for providing navigation assistance are focused on outdoor~\cite{velazquez2018-outdoor, heuten2008tactile-outdoor} and indoor environments~\cite{barontini2020integrating-indoor, huber2022vibrotactile-indoor}, particularly to aid individuals with limited vision. 
However, studies focusing on peripersonal navigation are relatively scarce and often limited to vertical search orientations~\cite{lehtinen2012dynamic-haptic,satpute2019-fingersight,dupont2020vibrotactile-verical} rather than accommodating diverse orientations. 
For example, FingerSight is a vibrotactile ring that enables users to navigate targets through a finger-mounted camera on a vertical wall~\cite{satpute2019-fingersight}. 
Similarly, PalmSight is a haptic device with a palm-mounted camera for object recognition and navigation~\cite{yu2016-palmsight}. 
While PalmSight can be used for any orientation, practical limitations exist due to its reliance on the hand constantly facing the scene, limiting its ability to capture horizontal surfaces effectively. 
Horizontal peripersonal navigation remains relatively underexplored due to challenges associated with obtaining a comprehensive top-down view with existing navigation devices~\cite{wei2022object-haptic,bahram-2012-caviar}. 
Given the diverse scenarios individuals encounter in peripersonal space navigation, there is a clear need to evaluate effective haptic guidance strategies that can accommodate diverse orientations.

To design an effective haptic guidance display for peripersonal navigation, there are multiple factors to consider.
Guidance approach, metaphor, and how to provide distance cues are the commonly investigated topics in the literature.
Depending on how tactors are positioned in haptic guidance displays, multiple guidance approaches could be utilized to send directional cues via tactons~\cite{brewster2004-tactons}. Tactons are a means of transmitting intricate messages using vibrotactile feedback by adjusting factors such as tactor numbers, vibration duration, amplitude, rhythm, and placement~\cite{brewster2004-tactons}. Multiple vibrations could be used in tactons either in succession or simultaneously to guide the hand in different positions. 
An example of providing simultaneous vibrations to convey oblique directions in confined spaces is PalmSight~\cite{yu2016-palmsight}, in which five tactors were positioned on the dorsal hand. 
These tactors would vibrate individually to indicate the cardinal directions (up, down, right, left), and vibrate simultaneously when conveying diagonal directions such as up-right. 
Given the possibility of diminished vibrotactile sensitivity in numerous anatomical locations when receiving simultaneous vibrations over short distances~\cite{shah-2019spatial, yeganeh2023discrimination, chen-2018effect, tajdari-2024sensitivity}, simultaneous vibrations may be less effective in haptic glove design than consecutive vibrations, despite providing the shortest path to target. Nevertheless, there is a lack of research comparing the effectiveness of guidance approaches in peripersonal navigation that utilize simultaneous vs. consecutive vibrations.

Guidance metaphor is another factor to consider, which refers to whether navigational vibrations are provided to move towards ("pull") or away ("push") from the vibrations.
The pull metaphor was considered more intuitive across many studies~\cite{lehtinen2012dynamic-haptic, jansen2004vibrotactile-push-pull,gunther2018-tactileglove}, by showing lower task completion time, lower cognitive workload, and also perceived as being more natural compared to the push metaphor~\cite{gunther2018-tactileglove}.
However, there is evidence supporting the effectiveness of the push metaphor in guiding hand motion. For instance, when teaching novice violin players proper bowing techniques, the push metaphor proved to be more effective~\cite{van-2010musicjacket}.
Additionally, there is no substantial evidence to support a potential interaction effect between the guidance metaphor and navigational orientations (vertical vs. horizontal). 
   
   Distance cues in navigational guidance are commonly conveyed through variations in vibration intensity~\cite{ wei2022object-haptic, lehtinen2012dynamic-haptic, froese2011-enactivetorch-et, de2023grasping}, frequency~\cite{gunther2018-tactileglove, pielot2010-pocketnavigator}, or duration~\cite{pielot2010-pocketnavigator, shih2018-dlwv2}. 
   Various approaches exist for representing distance to the target.
   For instance, vibration intensity may increase~\cite{wei2022object-haptic, froese2011-enactivetorch-et, de2023grasping} or decrease~\cite{lehtinen2012dynamic-haptic} as the target is approached.
   Intensity changes can be implemented either linearly~\cite{wei2022object-haptic, rueschen-2023-vibrotactile, straub-2009distance} or discretely, with distinct vibrations when the target is in close proximity~\cite{katzschmann2018-safe}. While some research suggests that users more easily detect discrete changes in vibration intensity compared to gradual increases aligned with distance~\cite{katzschmann2018-safe}, there remains a lack of comprehensive studies on which method is most effective for enhancing target navigation across diverse peripersonal settings.
   

   Lastly, the influence of biological sex on vibration perception has produced mixed results in the literature~\cite{bikah-2008supracutaneous-gender, post1994perception, gescheider1994effects-gender}.
   For instance, females have demonstrated greater sensitivity to vibrations in the hand~\cite{gescheider1994effects-gender}, whereas no significant differences were found for the wrist,  upper arm~\cite{bikah-2008supracutaneous-gender}, or dorsal hand~\cite{tajdari-2024sensitivity}. 
   Despite these findings, no studies have examined whether sex interacts with the strategies outlined for vibrotactile navigation systems. Therefore, it is crucial to explore whether specific haptic guidance strategies, across various navigational orientations, exhibit differential effectiveness based on sex.
   
  We evaluated various haptic guidance strategies using a custom-designed haptic glove tailored to facilitate navigation within peripersonal space across vertical and horizontal layouts.
    Three key design factors--\textit{guidance approach}, \textit{guidance metaphor}, and \textit{distance cue}--were tested in combination with vertical and horizontal \textit{scene layouts}.
    The evaluation focused on guidance efficiency, measured through task completion time, hand trajectory distance, and percentage of hand trajectory spent in critical areas, as well as user feedback. 
    Additionally, the potential impact of sex on these outcomes was investigated.
  
\section{Methods}
\label{methods}

    \subsection{Participants}
    A convenience sample of 22 participants (11 male and 11 female) were recruited from the Virginia Tech community and completed the study. 
    The mean (SD) age of male and female participants were 22.2 (1.2) and 26.1 (4.9) years, respectively. 
    To be eligible for the study, participants had to be 18 years or older, had normal to corrected normal vision, and be right-handed. 
    The study protocol was approved by the Institutional Review Board (IRB\#: 22-694) at Virginia Tech. 
    Written informed consent was obtained from all participants prior to any data collection.
    Participants were compensated either in cash at a rate of US \$10 per hour or through course credit.

    \subsection{Overview of the Haptic Guidance System}
\begin{figure}[H]
    \centering
    \makebox[\textwidth][c]{
        \includegraphics[scale=0.65]{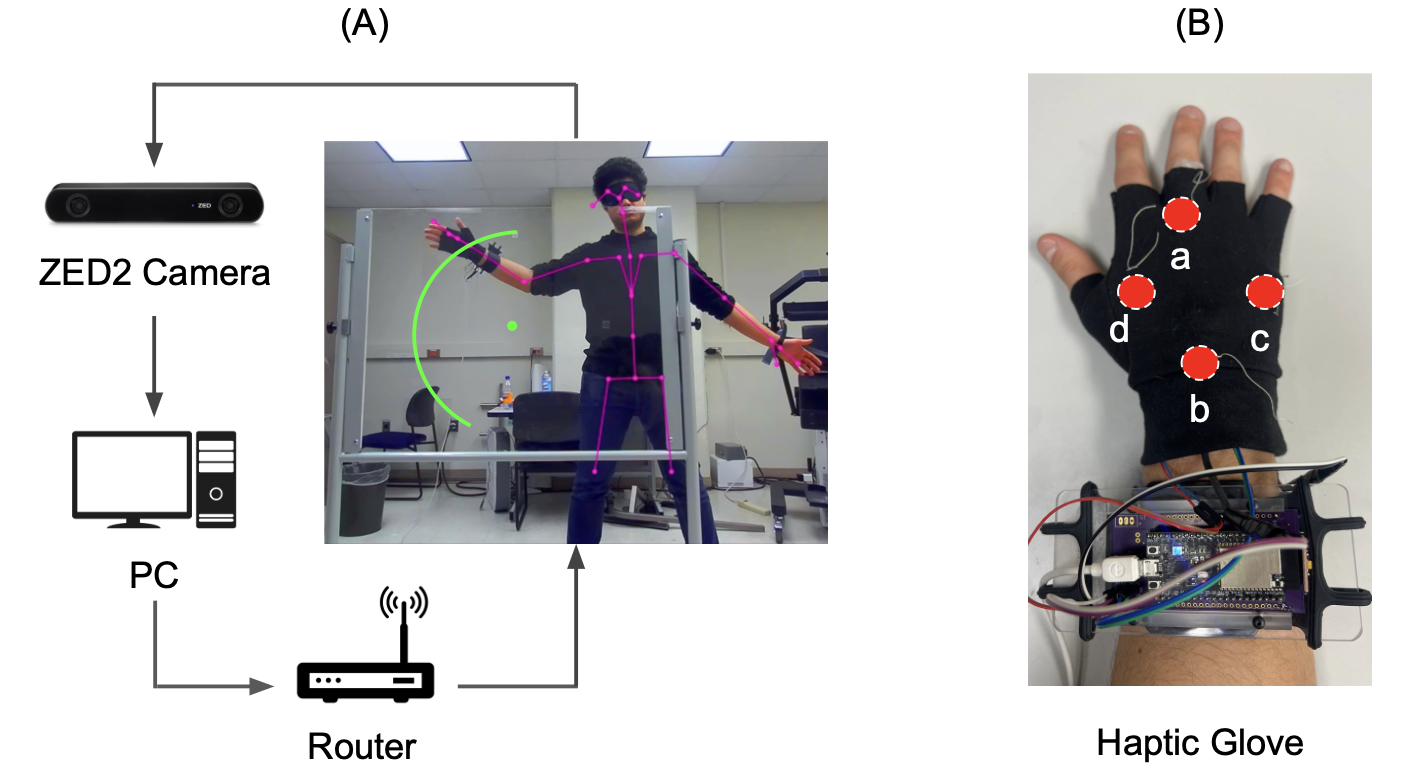}
    }
    \caption{(A): An overview of the system. A ZED2 camera captured the participant's video, which was then processed for joint tracking (indicated by pink dots representing the detected joint locations). The relative position of the hand from the target location was calculated, and the corresponding haptic feedback was subsequently transmitted to the glove via a wireless router. Green line = possible target locations. (B): The haptic glove used in this study, showing the four tactors (a, b, c, and d) positioned on the dorsal side of the hand.}
    \label{fig:overall-system}
\end{figure}

    We developed a custom haptic navigational guidance system (Figure~\ref{fig:overall-system}) capable of tracking a participant's hand location and transmitting real-time haptic feedback for navigational guidance. 
    Four eccentric rotating mass (ERM) coin tactors (Zhejiang Yuesui Electron Stock Co., Ltd, China; operating voltage = 2-3.6V) were affixed to specific locations on each participant's dorsal right hand using hypoallergenic medical tape (Figure~\ref{fig:overall-system}). These locations were: ($a$) the proximal phalanx of the middle finger, ($b$) between the Carpal bones on the wrist's dorsal side, ($c$) the middle of Metacarpal bones on the ulnar, and ($d$) radial (thumb) side.  
    Each motor location corresponded to the up, down, right, and left directions, respectively, and their locations were chosen based on our earlier work investigating sensitivity to haptic feedback~\cite{tajdari-2024sensitivity}.
    After attaching all four tactors to the participant's hand, the participant wore a fingerless glove made of lightweight elastic material. 
    We attached the tactors directly to the skin rather than to the glove. This allowed us to precisely control the motors' placement on the hand and to minimize potential interference from contact between the glove and the skin.
    
    A wristband with a flexible strap was used to secure an ESP32-S2 Arduino board (Espressif Systems, China) just above the participant's wrist, which controlled the tactors. 
    A router was used to establish the connection between the glove, the board, and a desktop computer that was connected to the camera tracking hand movements.
    The participant's hand movements were captured using a depth camera (ZED2, StereoLabs, USA) positioned in front of the participant.
    The software provided by the depth camera (the Body-34 skeleton model, StereoLabs, USA) detected the location of the participant's right hand (center of the palm) with a 30 Hz sampling rate. The data for each participant were stored in a .txt file upon completion of the trials, and can be accessed publicly via the following link: \url{https://github.com/mahdis98/Data-Haptic-Glove/tree/main}.

    \subsubsection{Target Generation}
\begin{figure}
            \centering
            \includegraphics[scale = 0.5]{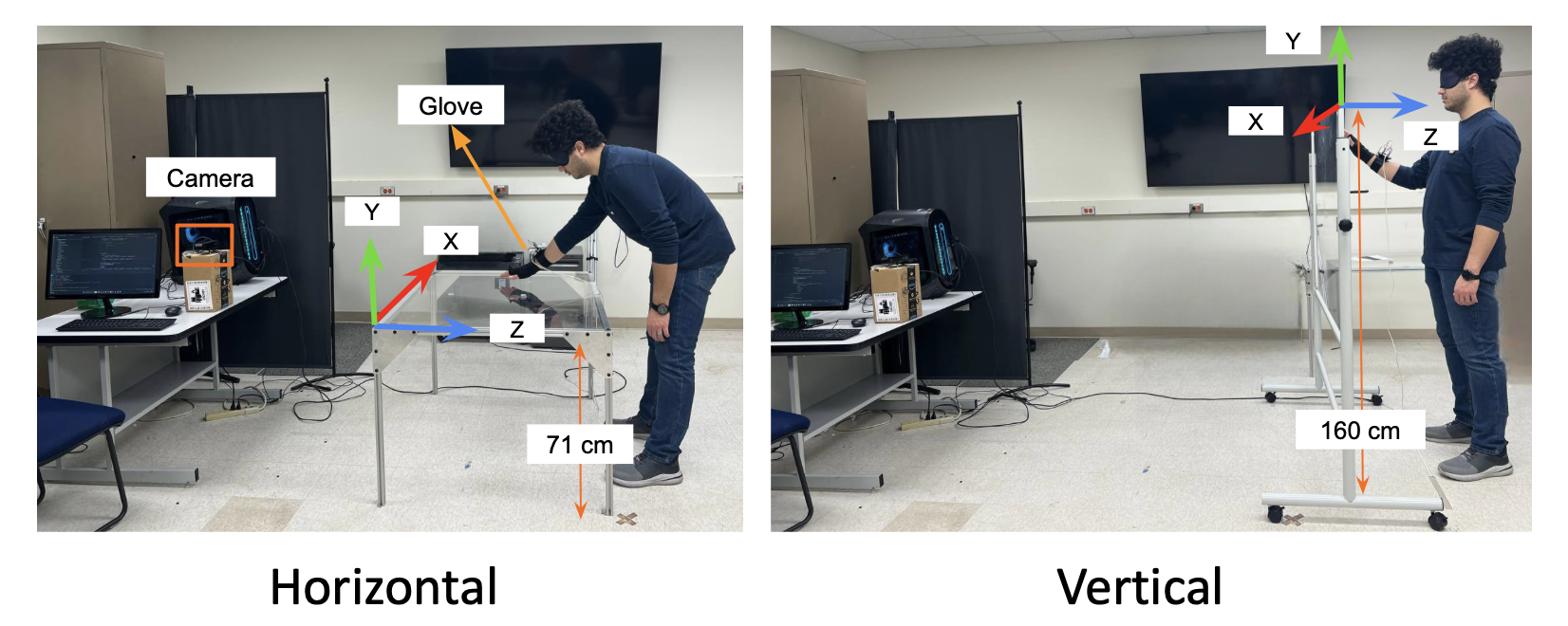}
            \caption{Horizontal and vertical scene layouts used in this study.}
            \label{fig:layout}
        \end{figure}

    In each trial, participants were asked to navigate to the target using their right hand as fast as possible from the center of the board (a green dot in Figure~\ref{fig:overall-system}).
    Targets were randomly generated each time, either on the horizontal plane (x-z axes) or on the vertical plane (x-y axes) depending on the layout condition (Figure~\ref{fig:layout}).
    These targets were virtual, thus there was no visual indication of the target locations on the planes. 
    All targets were positioned around the perimeter of a circle with a radius of 35 cm, where the center of the circle was located at the center of the board.
    The target distance from the center was kept the same to maintain the same travel distance in all trials, but the participants were not informed about this. 
    The targets were generated only within a partial section of the circle (0--150$^\circ$ or 12--5 o'clock range) to avoid a temporary potential loss of motion tracking caused by collisions between the torso and the right arm.
    Consecutive targets were generated with a minimum difference of 60$^\circ$. Software calibration was performed prior to the experiment to adjust the camera location relative to the board (see~\nameref{appendix} for more details). 

    \subsubsection{Haptic Feedback}
    To guide the participant's hand toward the target location, we used four tactors (labeled $a, $b$, c$, and $d$ in Figure~\ref{fig:overall-system}) according to the designated guidance approach, guidance metaphor, and distance cue condition for each trial. 
    Feedback was provided based on vectors derived from the four motor positions, with each direction represented by unit vectors, $\vec{v_a}, \vec{v_b}, \vec{v_c},$ and $\vec{v_d}$.
    The $a$--$b$ direction represented the vertical axis, with $a$ as the positive direction, while the $c$--$d$ direction represented the horizontal axis, with $c$ as the positive direction. 
    
    The hand rotation of the participants was tracked to deliver near-real-time haptic feedback based on the rotation angle of the hand ($\gamma$). 
    For example, if a participant's right hand tilted 90$^\circ$ clockwise, feedback was directed to the motors along the left-right axis defined by the $a$--$b$ vector. In trials using the two-tactor approach, where two motors vibrated to indicate diagonal directions, the intensity of each motor was adjusted according to the hand's tilt angle (see \nameref{appendix} for more details).

    \subsection{Experimental Procedures}
    Participants were provided a brief tutorial for different haptic feedback conditions, specifically how different guidance approaches, guidance metaphors, and distance cues will be used in the experimental trials. 
    They were also informed about the termination point of each trial, which involved all tactors generating a simultaneous buzzing for one second.
    After the tutorial, the tactors were attached to participants' hands using medical tape, and the board was positioned on their arm using a wristband. 
    
    For calibration of the system, participants were asked to hold their right palm facing the camera for five seconds at five different locations.
    The five locations were marked on the board used for the vertical layout setting and located on a perimeter of the target circle (with a 35-cm radius), each at 0, 90, 180, and 270$^\circ$ angles. Lastly, participants were asked to hold their palm at the center point of the circle (see \nameref{appendix} for a detailed explanation about the calibration procedure). 
    
    Before the trial, participants were blindfolded using an elastic eye mask to simulate impaired vision condition, and were given two test trials. 
    Each participant performed 48 trials (2 scene layouts $\times$ 2 guidance approaches $\times$ 2 guidance metaphors $\times$ 2 distance cues $\times$  3 repetitions). Scene layouts were presented in a counterbalanced order between participants, and guidance approach, guidance metaphor, and distance cue were randomized within the same scene layout condition. 
    After the completion of each trial, participants were asked to complete a short survey (Table~\ref{table:appendixA}) on a 7-point Likert scale verbally. 
    Participants were allowed to ask for a break or withdraw from this study at any point. After completing the trials, participants were asked to complete a set of open-ended questions (Table~\ref{table:appendixB-open}) regarding their overall experience throughout the experiment. 

\begin{table}[H]
    \centering
    \caption{Usability questions using a 7-point Likert scale (1: strongly disagree, 7: strongly agree). 
    }
    \label{table:appendixA}
    \resizebox{\textwidth}{!}{ 
    \begin{tabular}{l}
        \midrule
        \textbf{Q1.} I could easily navigate targets. \\
        \textbf{Q2.} Haptic cues acted based on my navigational instincts. \\
        \textbf{Q3.} I didn't feel insecure, discouraged, irritated, stressed, or annoyed while using the navigation system. \\
        \bottomrule
    \end{tabular}
}
\end{table}


\begin{table}[H]
    \centering
    \caption{The open-ended post-study questions.}
    \label{table:appendixB-open}
    \resizebox{\textwidth}{!}{ 
        \begin{tabular}{l}
            \midrule
            \textbf{Q1.} What did you like most about using the glove? \\
            \textbf{Q2.} What did you like least about using the glove? \\
            \textbf{Q3.} If you could change anything about the glove, what would you change? \\
            \textbf{Q4.} Do you think this haptic glove would help visually impaired individuals with navigating their belongings in an indoor \\
            environment (e.g., their bedroom)? Why? \\
            \bottomrule
        \end{tabular}
    }
\end{table}

    Four independent variables manipulated in the study are as follows:

    \begin{itemize}
        \item \textbf{Scene Layout (Vertical vs. Horizontal)}: 
        Custom see-through boards made of plexiglass (120 x 91 cm) were mounted on metal frames to support both horizontal and vertical layout configurations (see Figure~\ref{fig:layout}).
         
        \item \textbf{Guidance Approach (Two-Tactor vs. Worst-Axis First)}: Based on prior research~\cite{satpute2019-fingersight}, two guidance approaches were implemented. 
        In the two-tactor approach, two tactors vibrate simultaneously to guide participants along the most direct (oblique) path to the target.
        In contrast, the worst-axis first approach conveys direction using one tactor at a time, starting with the axis farthest from the target (Figure~\ref{fig:approach-mataphor}). 
        For example, if the hand is at (0,0) and the target is at (5,1), the horizontal axis is corrected first with a vibrating tactor, followed by vertical correction as needed. 
        This sequential approach divides movement into separate steps for each axis, with only one tactor active at a time.

         \begin{figure}
            \centering
            \includegraphics[scale = 0.55]{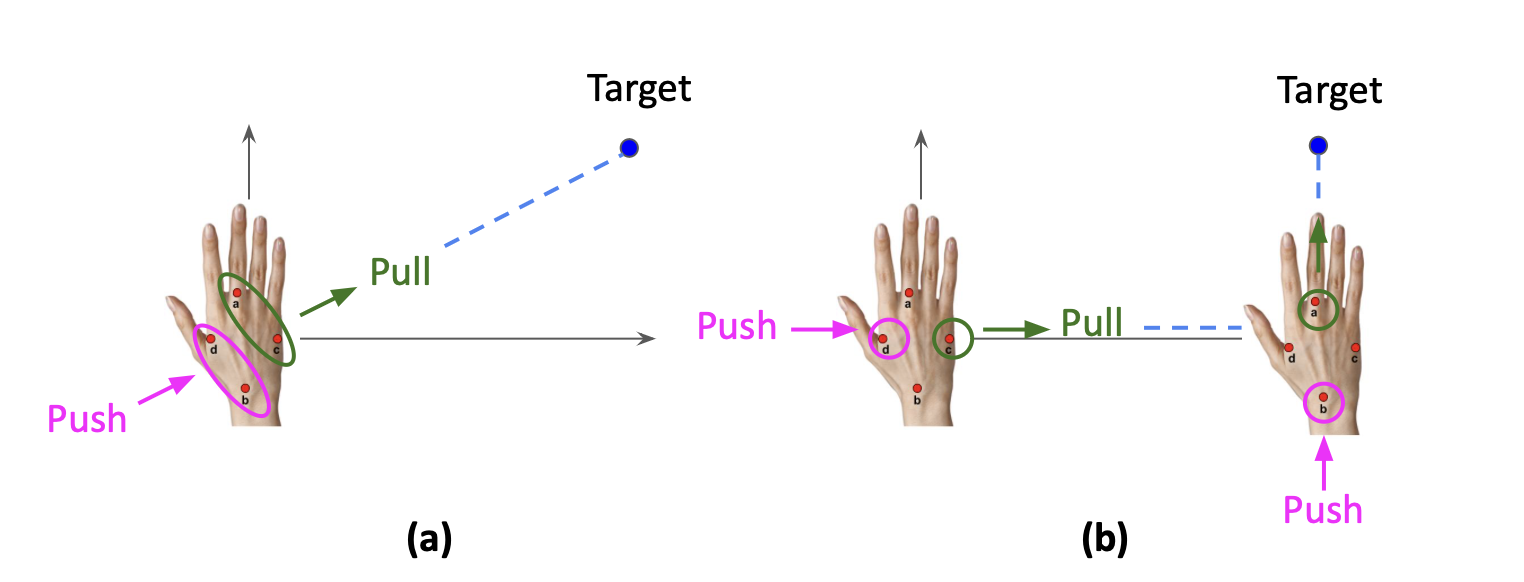}
            \caption{Guidance approaches and metaphors used in this study: (a) Two-tactor guidance approach, and (b) Worst-axis first approach. Tactors circled in green indicate "Pull" mode vibrations, while those circled in pink indicate "Push" mode. The blue dashed line represents the ideal path that participants should follow in each condition to reach the target.}

            \label{fig:approach-mataphor}
        \end{figure}

        \item \textbf{Guidance Metaphor (Push vs. Pull)}: In the push metaphor, participants were instructed to move in the opposite direction of the vibration, simulating the sensation of being "pushed" from the vibration point to guide their movement. In contrast, during the pull mode, they were expected to move in the same direction as the vibration, simulating the sensation of having their hand "pulled" in that direction (see Figure~\ref{fig:approach-mataphor}). 
        
        \item \textbf{Intensity Mode (Linear vs. Zone)}: Tactor intensity was adjusted to convey distance information using two modes. In the linear mode, the intensity of the active tactors increased proportionally as the participant's hand approached the target. In the zone mode, however, intensity remained constant until the hand entered a predefined "zone" surrounding the target, set as a 7 cm-radius circle. Upon entering this zone, the intensity sharply increased and stayed consistent as long as the participant remained within the zone (see Figure~\ref{fig:intensity}). 
        \begin{figure}
            \centering
            \includegraphics[scale = 0.35]{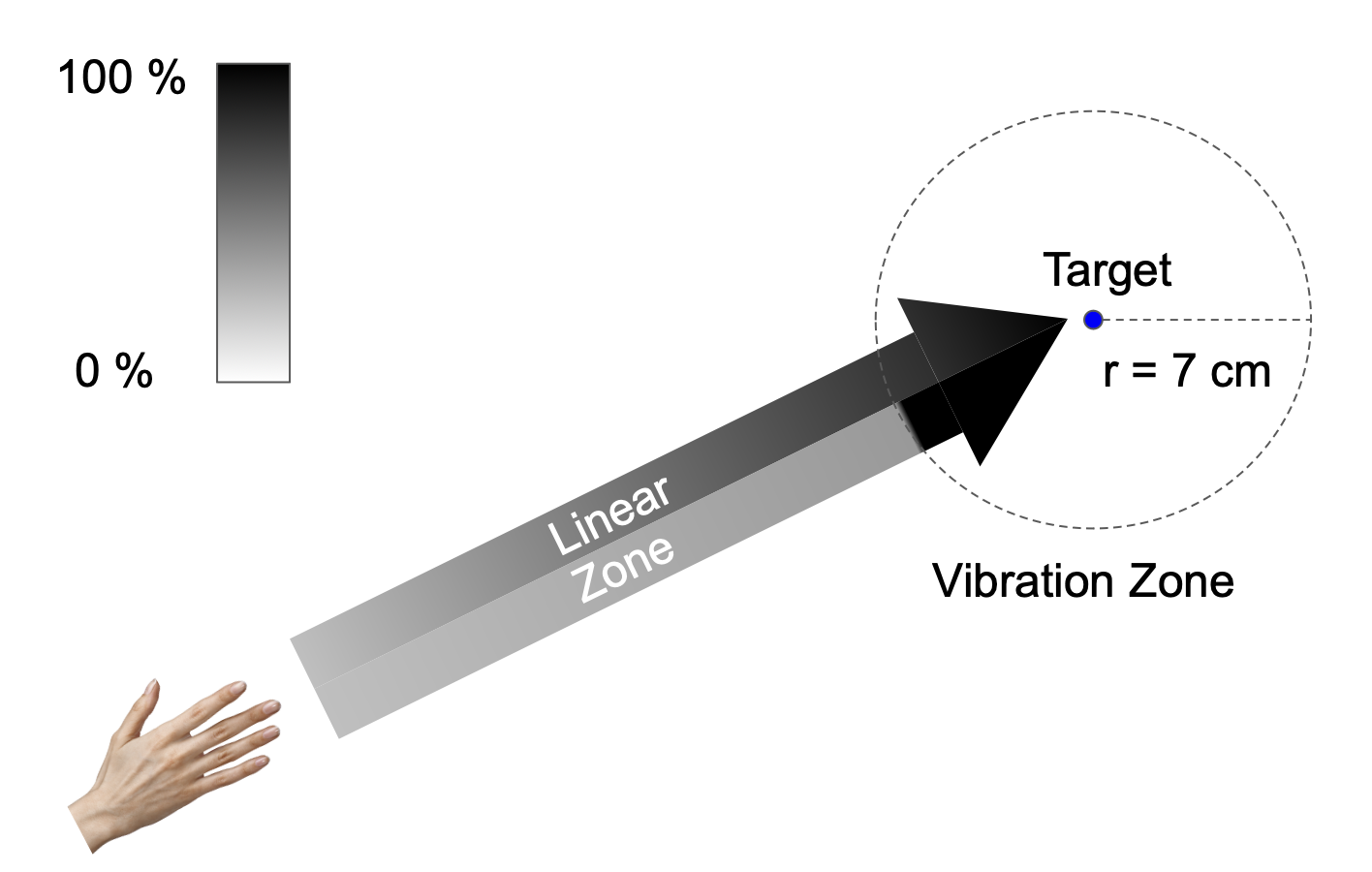}
            \caption{Intensity modes to convey distance cues: darker shades indicate stronger vibration levels compared to lighter shades.
            }
            \label{fig:intensity}
        \end{figure}

    \end{itemize}

   \begin{flushleft}

    \subsection{Dependent Variables}
    Three performance measures were calculated and used as dependent variables:

\begin{itemize}
    \item \textbf{Task Completion Time (seconds):} The time each participant required to reach the virtual target in each trial, starting from the center of the board.
    \item \textbf{Hand Trajectory Distance (cm):}  The total distance covered by each participant's hand during a trial. 
    \item \textbf{Percentage of Hand Trajectory in Critical Area (\%):} This metric evaluated how effectively vibrations guided participants to the target. A critical region, represented by a dashed circle (Figure~\ref{fig:zone-radius}), was defined around the optimal path to the target with a 20\% margin allowance. The percentage was calculated as the proportion of the participant’s hand trajectory that stayed within this critical region relative to the total hand trajectory length during the trial. 
\end{itemize}
    
     \begin{figure}[h]
            \centering
            \includegraphics[scale = 0.3]{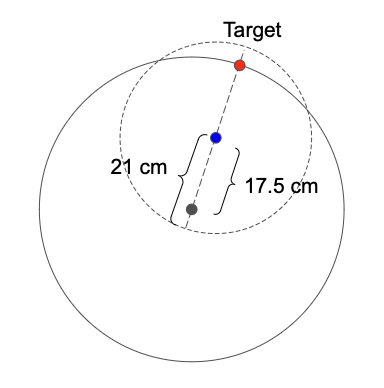}
            \caption{A dashed circle represents the critical region used to calculate the proportion of the hand trajectory near optimal paths relative to the total distance traveled. The center of this critical region (shown in blue) is positioned midway between the center of the main circle (gray) and the target (red). The radius of the critical region is set to 21 cm, providing a 20\% margin to accommodate variations in hand movement.}
            \label{fig:zone-radius}
        \end{figure}

    Three usability questions (Table~\ref{table:appendixA}), representing the ease of navigation (Q1), cue alignment (Q2), and absence of frustration (Q3) were also used as dependent variables.   
        
   \subsection{Statistical Data Analysis}
    \justifying
   Mixed effects models were used to investigate the effects of scene layout, guidance approach, guidance metaphor, intensity mode, and sex on task completion time, hand trajectory distance, percentage in critical area, and quantitative post-study questions. 
   The restricted maximum likelihood (REML) method was used for the mixed effects model and participants were considered as a random effect. 
   We log-transformed task completion time and hand trajectory distance before the statistical tests, as the initial residual distribution was not close to a normal distribution. 
   The study examined up to two-way interactions of layout, metaphor, approach, and intensity, and utilized the Tukey-HSD test to identify significantly  different pairs where a significant interaction effect was observed.  All tests were conducted using JMP Pro (v16.0.0, SAS, NC, USA), and they were concluded at $\alpha = 0.05$.

    \end{flushleft}

\section{Results}
\label{results}
Mixed effects model results for task completion time, hand trajectory distance, and percentage of time spent in critical area are summarized in Table~\ref{tab:effects_objective}.

\subsection{Task Completion Time}

\begin{figure}
            \centering
            \includegraphics[width=\textwidth]{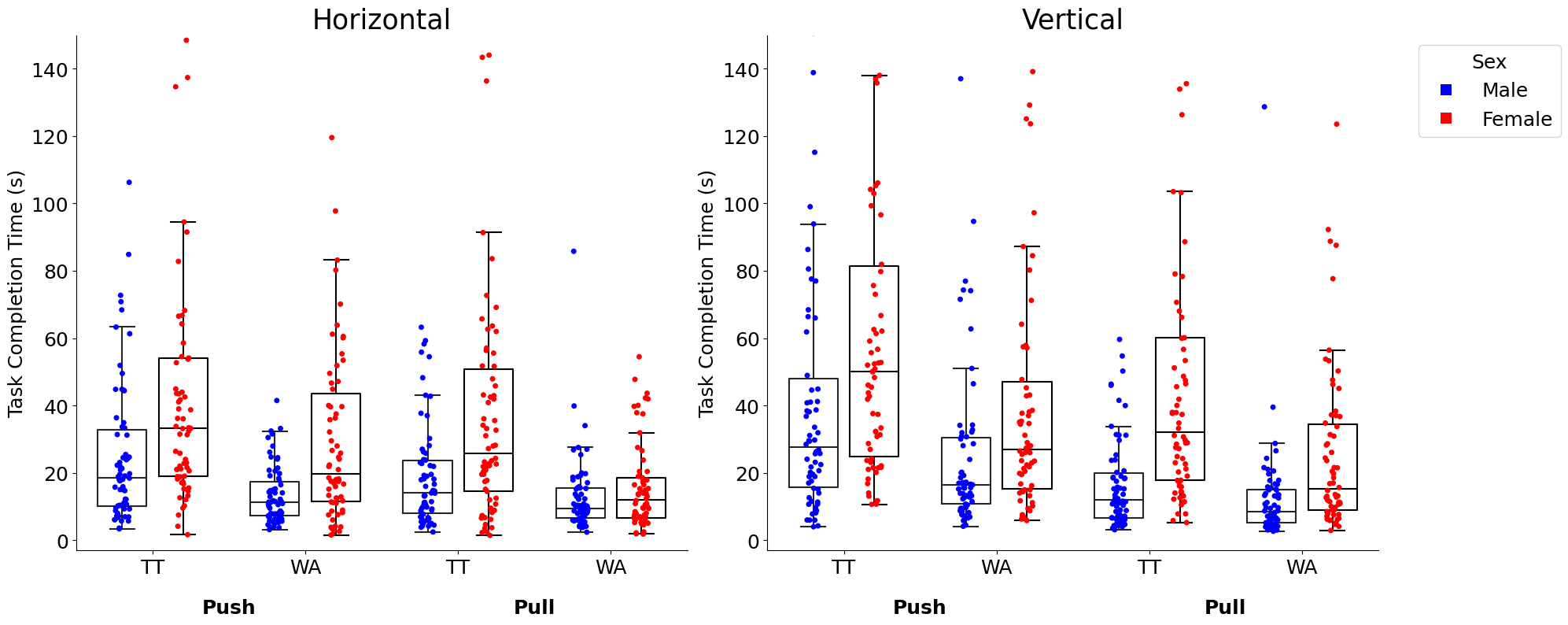}
            \caption{Task completion time (s) across different \textit{layouts}, \textit{approaches}, and \textit{metaphors}, stratified by sex (WA: Worst-axis first, TT: Two-tactor).}
            \label{fig:task-completion}
        \end{figure}

        Significant main effects of layout, approach, metaphor, and sex, as well as an interaction effect between layout and metaphor were found for task completion time (Figure~\ref{fig:task-completion}).
        Task completion time was significantly shorter with the horizontal layout (11.0 seconds faster than the vertical layout), the worst-axis first approach (16.9 seconds faster than the two-tactor approach), the pull metaphor (14.9 seconds faster than the push metaphor), and among male participants (20.5 seconds faster than female participants).

        The interaction effect between layout and metaphor revealed that the vertical-push condition resulted in significantly slower task completion times compared to all other layout and metaphor combinations: 21.1 seconds slower than vertical-pull ($p < .0001$), 26.0 seconds slower than horizontal-pull ($p < .0001$), and 17.2 seconds slower than horizontal-push ($p < .0001$). 
        Within the horizontal layout, task completion time was also significantly slower (by 8.8 seconds) when the push metaphor was used ($p = 0.04$).

\begin{table}
    \centering
    
    \caption{Summary of results from the mixed effects model for the effects of Layout, Approach, Metaphor, Intensity, and Sex on task completion time, hand trajectory distance, and percentage of hand trajectory in critical area. Significant effects are in bold font.}
    \vspace{5pt}
    \resizebox{\textwidth}{!}{
        \begin{tabular}{c@{\hskip 4em}cc@{\hskip 4em}cc@{\hskip 4em}cc}
            \toprule
            \multirow{2}{*}{\centering \vspace{0.005ex} Effect \vspace{0.005ex}} & \multicolumn{2}{c@{\hskip 4em}}{Time} & \multicolumn{2}{c@{\hskip 4em}}{Distance} &  \multicolumn{2}{c@{\hskip 4em}}{\% of Trajectory in Area} \\
            \cmidrule{2-7}
            & F-Ratio & Prob>F & F-Ratio & Prob>F & F-Ratio & Prob>F \\
            \midrule
            Layout & \textbf{33.37}  & \textbf{<.0001} & \textbf{37.45} & \textbf{<.0001} & \textbf{29.14}& \textbf{<.0001}\\
            Approach & \textbf{108.25} & \textbf{<.0001} & \textbf{77.64} & \textbf{<.0001} & \textbf{72.72} & \textbf{<.0001}\\
            Metaphor & \textbf{87.28} & \textbf{<.0001} & \textbf{59.41} & \textbf{<.0001} & 1.96 &  .16\\
            Intensity & .08 & .78 & .06 & .80 & 1.34 & .25\\
            Sex & \textbf{13.15} & \textbf{.002} & \textbf{17.85} & \textbf{.0004} & \textbf{7.57} & \textbf{.01}\\
            Layout $\times$ Metaphor & \textbf{10.08} & \textbf{.002} & \textbf{12.43} & \textbf{.0004} & \textbf{12.7} & \textbf{.0004}\\
            Layout $\times$ Approach & .05 &  .83 & .02 & .90 & \textbf{6.07} & \textbf{.01}\\
            Approach $\times$ Metaphor & .09 & .77 & .03& .87& 2.46 & .12\\
            Layout $\times$ Intensity & .16 & .69 & .002 & .97 & 2.03 & .15\\
            Approach $\times$ Intensity & .24 & .63 & .06 & .81 & .38 & .54\\
            Metaphor $\times$ Intensity & .93 & .33 & .006 & .94 & .002 & .97\\
            
            \bottomrule
        \end{tabular}
    }
    \label{tab:effects_objective}
\end{table}

        \subsection{Hand Trajectory Distance}
\begin{figure}
            \centering
            \includegraphics[width=\linewidth]{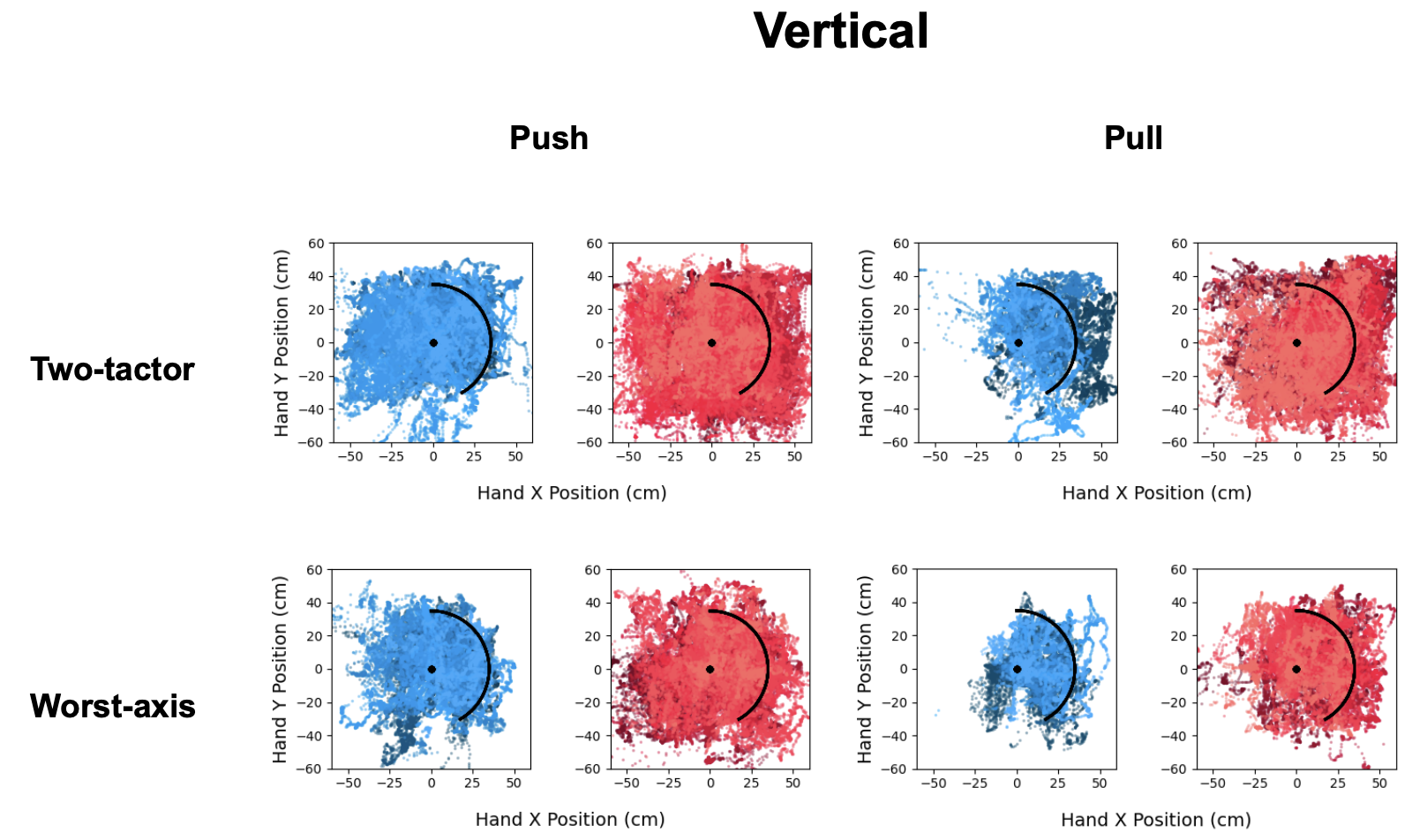}
            \\[2.5em]
            \includegraphics[width=\linewidth]{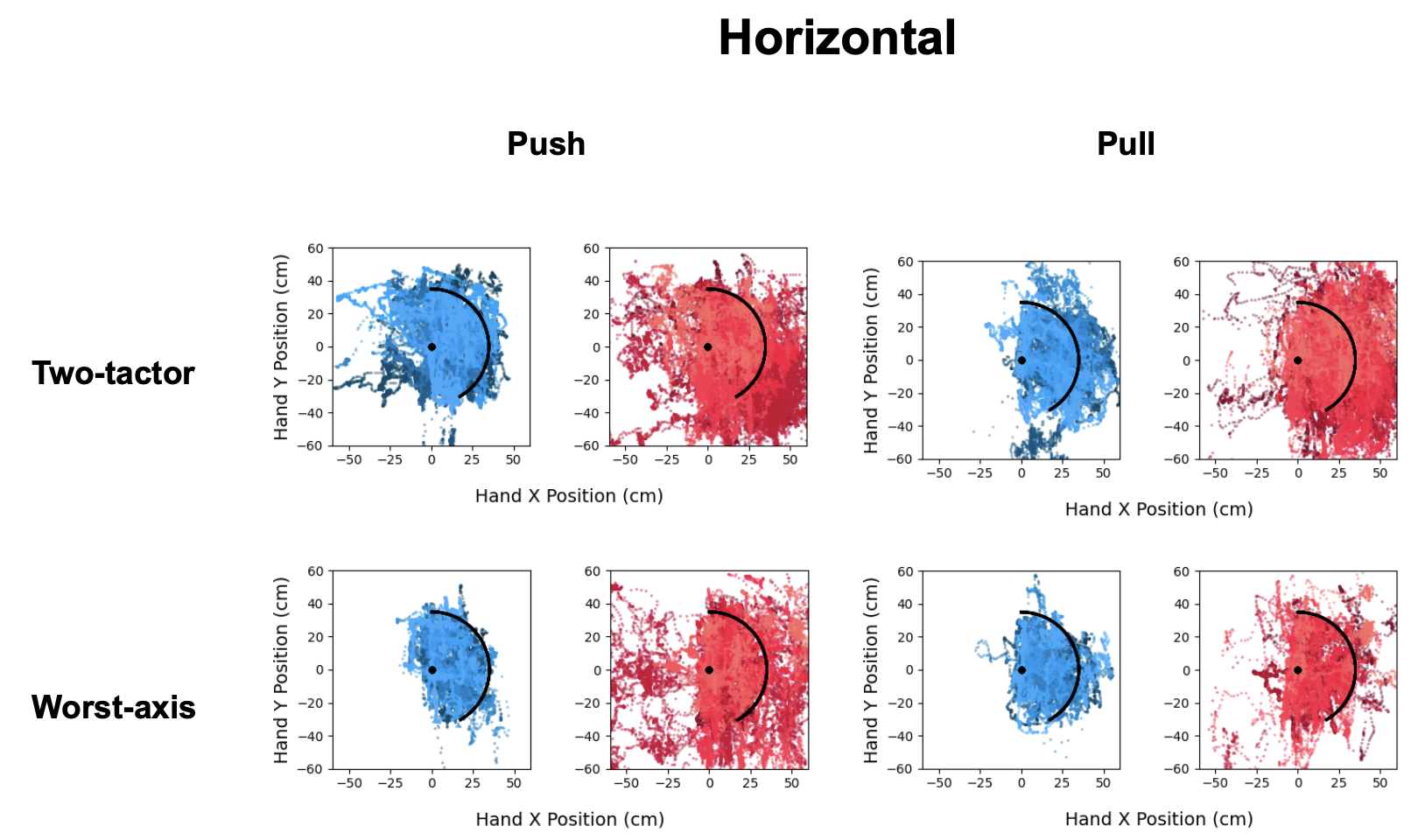}
            \caption{Hand trajectories from all participants across different layout, metaphor, and approach conditions, shown separately for male (blue) and female (red) participants. The black circular line indicates potential target locations.}
            \label{fig:trajectories}
        \end{figure}

        Figure~\ref{fig:trajectories} shows the hand trajectories from all participants across different layout, metaphor, and approach conditions. 
         
\begin{figure}
                \centering
                \includegraphics[width=\textwidth]{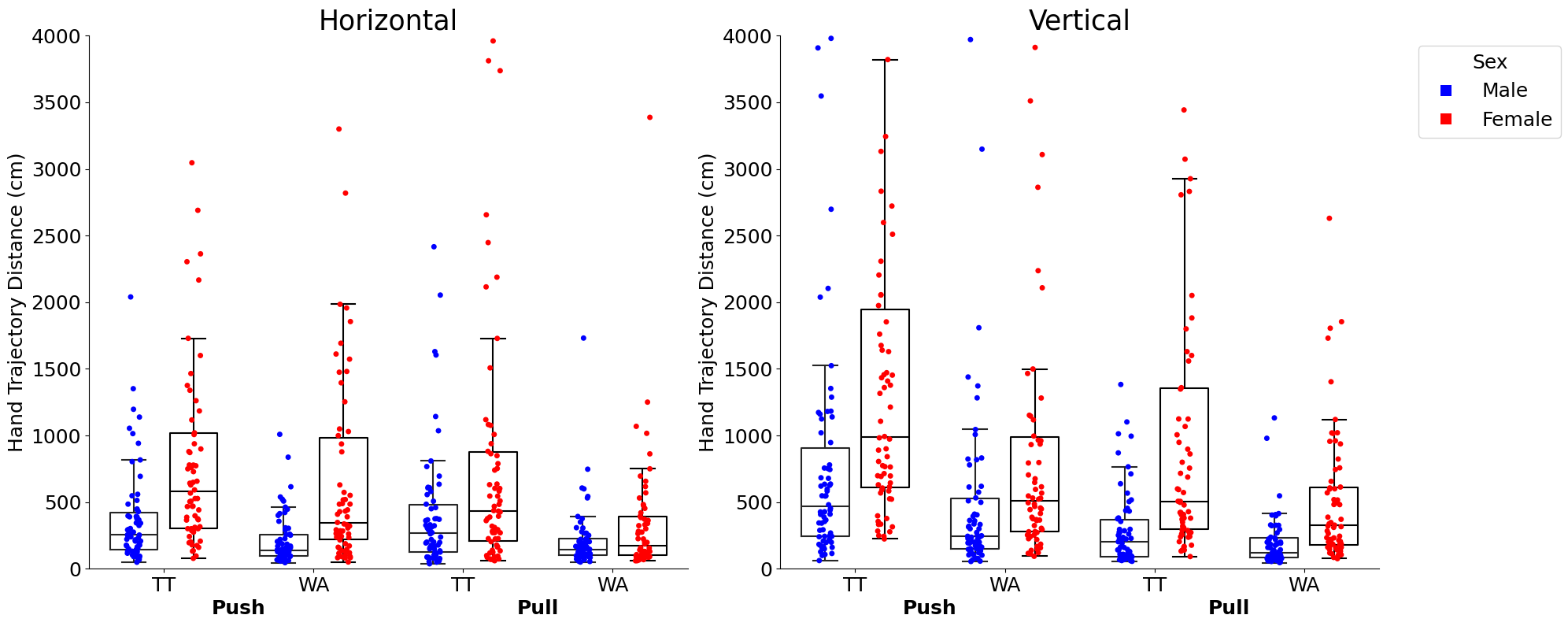}
                \caption{Hand trajectory distance across different \textit{layouts}, \textit{approaches}, and \textit{metaphors}, stratified by sex (WA: Worst-axis first, TT: Two-tactor).
                }
                \label{fig:distance_all}
            \end{figure}

        Similar to task completion time, significant main effects of layout, approach, metaphor, and sex, as well as an interaction effect between layout and metaphor were found for hand trajectory distance (Figure~\ref{fig:distance_all}). 
        
        Participants moved a significantly shorter distance to reach to their targets in the horizontal layout (432.3 cm shorter than the vertical layout), the worst-axis first approach (354.7 cm shorter than the two-tactor approach), and the pull metaphor (170.4 cm shorter than the push). 
        Moreover, male participants outperformed females by traveling a significantly shorter distance, with a mean difference of 314.9 cm (Figure~\ref{fig:distance_all}). 
        
        The interaction effect showed that the vertical-push condition resulted in significantly shorter trajectory distance compared to all other layout and metaphor combinations: 240.7 cm shorter than vertical-pull ($p < .0001$), 603.2 cm shorter than horizontal-pull ($p < .0001$), 502.3 cm shorter than horizontal-push ($p < .0001$). 
        Within the horizontal layout, hand trajectory distance was also significantly shorter (by 100.9 cm) when the push metaphor was used ($p = .02$).

        \subsection{Percentage of Hand Trajectory in Critical Area}
\begin{figure}[h]
            \centering
            \includegraphics[width=\textwidth]{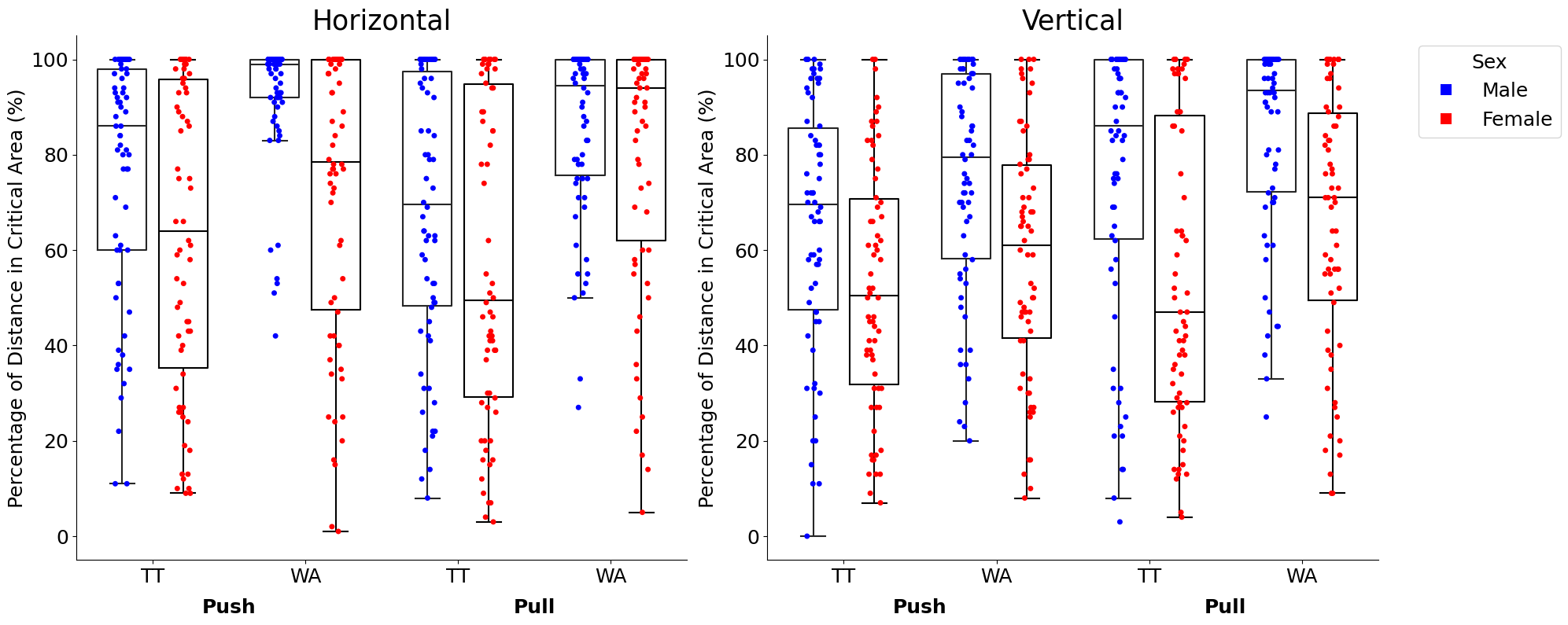}
            \caption{Percentage of distance spent in the critical area around the target (see Figure~\ref{fig:zone-radius}) across different \textit{layouts}, \textit{approaches}, and \textit{metaphors}, stratified by sex (WA: Worst-axis first, TT: Two-tactor).}
            \label{fig:percentage_time_all}
        \end{figure}

        Significant main effects of layout, approach, and sex, as well as significant interaction effects between layout $\times$ metaphor and layout $\times$ approach were found for percentage of hand trajectory in critical area (Figure~\ref{fig:percentage_time_all}).
        Participants' hands stayed in the critical area significantly more with the horizontal layout (8.0\% more than the vertical layout),  the worst-axis first approach (12.6\% more than the two-tactor approach), and among male participants (14.7\% more than female participants). 

        

        The interaction effect showed that participants spent significantly less proportion of hand trajectory in the critical area in the vertical-push condition compared to all other layout and metaphor combinations: 20.56\% less than horizontal-push ($p < .0001$), 10.04\% less than horizontal-pull ($p < .0001$), and 7.33\% less than vertical-pull ($p = .003$). A significant difference was also found between the horizontal-push and vertical-pull conditions, with horizontal-push resulting in 5.92\% less trajectory in the critical area ($p = .02$).
        

        Moreover, the interaction effect showed that the horizontal-worst-axis condition resulted in significantly more trajectory in the critical area compared to all other layout and approach combinations: 20.56\% less than vertical-two-tactor ($p < .0001$), 16.22\% less than horizontal-two-tactor ($p < .0001$), and 11.61\% less than vertical-worst-axis ($p < .0001$). A significant difference was also observed between the vertical-worst-axis and vertical-two-tactor conditions, with vertical-worst-axis resulting in 8.95\% more trajectory in the critical area ($p = .0001$).

       \justifying
        \subsection{Usability}
\begin{figure}
            \makebox[\textwidth][c]{
        \includegraphics[scale=0.5]{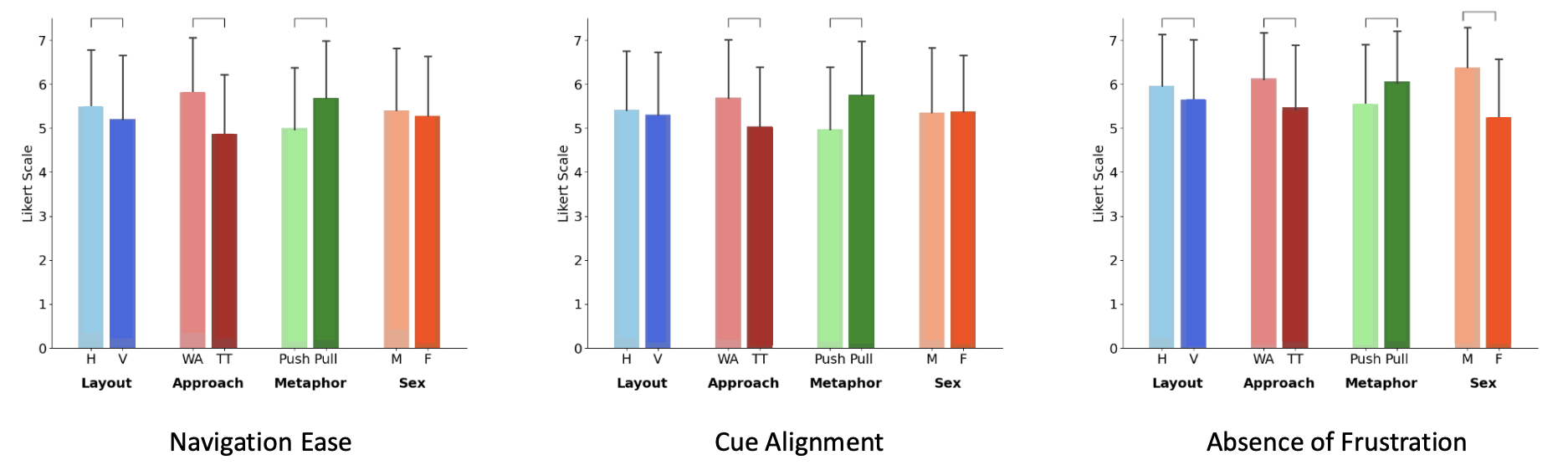}}
            \caption{Responses on the three usability questions on a 7-point Likert scale (H: Horizontal, V: Vertical, WA: Worst-axis first, TT: Two-tactor, M: Male, F: Female). The significantly different pairs are connected with a bar.}
        \label{fig:appendixA-main}
        \end{figure}

        Mixed effects model results for usability questions are summarized in Table~\ref{tab:Q}. 
        Significant main effects of layout, approach, and metaphor were found for navigational ease (Q1). 
        Participants perceived horizontal layout, worst-axis first approach, and pull metaphor significantly easier to navigate. 
        Significant main effects of approach and metaphor were found for cue alignment with their instincts (Q2). 
        Participants reported worst-axis first approach and pull metaphor as more aligned with their navigational instincts, compared to the alternative conditions. 
        Lastly, there were significant main effects of layout, approach, metaphor, and sex for negative user perception (Q3). 
        Horizontal layout, worst-axis first approach, and pull metaphor were perceived more positively. 
        Male participants also rated the system more positively compared to the female participants (Figure~\ref{fig:appendixA-main}). 

     \begin{table}[t]
    \centering
    \caption{Summary of results from the mixed effects model for the effects of Layout, Approach, Metaphor, and Sex on usability. Significant effects are highlighted using bold font.}
    \vspace{5pt}
    \resizebox{\textwidth}{!}{ 
    \begin{tabular}{c@{\hskip 4em}cc@{\hskip 4em}cc@{\hskip 4em}cc}
        \toprule
        \multirow{2}{*}{\centering Effect} & \multicolumn{2}{c@{\hskip 4em}}{Navigation Ease} & \multicolumn{2}{c@{\hskip 4em}}{Cue Alignment} &  \multicolumn{2}{c@{\hskip 4em}}{Absence of Frustration} \\
        \cmidrule{2-7}
        & F-Ratio & Prob>F & F-Ratio & Prob>F & F-Ratio & Prob>F \\
        \midrule
        Layout & \textbf{7.13} & \textbf{.008} & 1.06 & .3 & \textbf{11.77} & \textbf{.0007} \\
        Approach & \textbf{76.5} & \textbf{<.0001} & \textbf{37.64} & \textbf{<.0001} & \textbf{51.48} & \textbf{<.0001} \\
        Metaphor & \textbf{39.37} & \textbf{<.0001} & \textbf{53.36} & \textbf{<.0001} & \textbf{30.83} & \textbf{<.0001} \\
        Intensity & 1.3 & .25 & 0.12 & .72 & 1.13 & .29 \\
        Sex & .07 & .79 & 0.27 & .6 & \textbf{14.91} & \textbf{.0007} \\
        Layout $\times$ Metaphor & 2.41 & .12 & .16 & .7 & .19 & .66 \\
        Layout $\times$ Approach & 0.51 & .48 & .86 & .35 & 0.004 & .95 \\
        Approach $\times$ Metaphor & .16 & .69 & .0008 & .98 & .19 & .66 \\
        Layout $\times$ Intensity & .51 & .48 & .0008 & .99 & .004 & .95\\
        Approach $\times$ Intensity & .0001 & .99 & .67 & .41 & .004 & .95\\
        Metaphor $\times$ Intensity & .37 & .54 & .25 & .62 & .31 & .58\\
        \bottomrule
    \end{tabular}
}

    \label{tab:Q}
\end{table}

        In the semi-structured open-ended questions conducted at the study's conclusion (see Table~\ref{table:appendixB-open}), participants shared their preferences and perceptions regarding the haptic guidance strategies and their overall experience with the haptic-assisted peripersonal navigation system. 
        Participants generally reported a positive experience, emphasizing the system’s usefulness, intuitive navigation, and thoughtful placement of tactors. They appreciated the non-restrictive fit of the glove but noted areas for improvement to enhance comfort and usability. Some participants reported discomfort due to the weight of the wrist-mounted board compartment, suggesting reducing its weight or relocating it higher up, such as over the shoulder. Others recommended adjusting the spacing and positioning of the tactors, particularly near the wrist, to reduce sensory confusion. Additional feedback included decreasing the intensity of vibrations to prevent discomfort during prolonged use and customizing haptic feedback strategies to suit individual preferences and the interacting surface.
        
        Navigation and target recognition were described as intuitive, particularly when using the pull mode. One participant particularly mentioned: \textit{"When it was one tactor and pull, it helped me go directly to the target; it was instinctive and I didn't have to think more"}. 
Many preferred receiving the vibration from one tactor at a time, as this minimized feelings of overwhelm and confusion. Additionally, many participants found it challenging to distinguish which tactors were vibrating in the two-tactor mode, particularly in the vertical layout. The close proximity of the tactors on the dorsal hand was also a common issue, with many participants mistaking the bottom tactor near the wrist for those on the sides.  A few participants also recommended customizing haptic feedback strategies based on individual preferences and the interacting surface, along with incorporating warning signals for nearby dangerous objects (e.g., knives in the kitchen).


        Regarding the layout, the design choice between horizontal and vertical configurations had a notable impact on navigation ease. The horizontal layout was found to enhance navigation compared to the vertical setting. One participant's feedback emphasized the intuitive design of the horizontal layout, which made it easier to track hand positions on the table. As the participant noted, \textit{"I thought it was very intuitive, I was able to tell where my hand was on the table".} This suggests that the horizontal layout offers a more natural and accessible navigation experience.

        The distance cues did not significantly impact the results, with only a few participants explicitly noting that increased intensity aided in target navigation. One participant specifically commented, \textit{"The haptics help a lot as soon as you get close to the target, as they [the tactors] vibrate harder".} However, there were no other references to the distance cues being particularly useful for locating the targets.

        Overall, participants were optimistic about the potential of the haptic glove to assist visually impaired individuals or in vision restricted conditions (e.g., dark or foggy environment) in peripersonal navigation and object localization. Several participants noted that the glove provided accurate directional guidance, in contrast to auditory feedback, which could be unreliable in noisy environments. Additionally, participants reported being able to quickly learn the system after just two training trials.
        
\section{Discussion}
\label{discussion}

The present work proposes a haptic glove for peripersonal target navigation of vertical and horizontal 2-D environments, employing and comparing two guidance approaches, guidance metaphors, and tactor intensity modes. 
The subsequent sections delve into the study's findings, providing valuable insights for the design of vibrotactile navigation devices tailored for both horizontal and vertical peripersonal spaces. Concluding this section, the study's limitations are acknowledged, and future research directions in this field are outlined.

\subsection{Scene Layout}
Participants demonstrated significantly better performance in the horizontal layout compared to the vertical layout, as evidenced by reduced task completion times and shorter distances traveled. 
This finding aligns with the participants' subjective feedback, which highlighted the horizontal layout's ease of navigation and reduced frustration levels. 
A plausible explanation for these findings is that horizontal layouts, such as tabletops, offer a natural and ergonomic surface that supports the weight of the hand and arm. 
This support enables more comfortable wrist and arm positions, reducing physical strain. 
In contrast, vertical layouts necessitate non-neutral shoulder, arm, and wrist positions, increasing physical efforts and diminishing the precision of the movement.
The additional strain, particularly from increased shoulder flexion angle~\cite{antony-2010effects} likely contributed to slower task completion times and reduced overall navigational efficiency. 

Participants’ behavior further supports these findings. In the horizontal layout, they remained relatively stationary, relying primarily on arm and hand movements to navigate. In contrast, the vertical layout prompted participants to step sideways or walk around to explore different sections of the board. This additional whole-body movement likely contributed to longer task completion times and increased distances traveled.

Interestingly, participants devoted significantly more time in the horizontal layout to closely examining areas near the target during each trial. This suggests they could more easily discern the general direction of the target, allowing for more precise adjustments. Conversely, in the vertical layout, participants spent less time in close proximity to targets but transitioned more quickly from identifying the general direction to locating the target, reflecting a trade-off between speed and precision influenced by the layout configuration.

In real-world scenarios, users typically have limited control over the placement of targets within their environment. Our findings offer valuable insights for optimizing haptic navigation systems based on the orientation of the search task. For instance, when a target is located beyond a participant's immediate reach envelope, it may be beneficial to employ a two-phase haptic feedback strategy.
Initially, a distinct type of vibrotactile signal could guide users to move closer to the target by taking a few steps. Once within closer proximity, more precise haptic feedback can be provided to assist with fine-grained navigation. This approach is particularly effective for vertical layouts, such as bookshelves or kitchen cabinets, where reaching the target may require both gross body movement and precise hand coordination. Such adaptive feedback mechanisms could significantly enhance the efficiency and user experience in diverse spatial configurations.

\subsection{Guidance Approach}

The two-tactor guidance approach resulted in significantly longer task completion times compared to the worst-axis first approach. Participants found the two-tactor method less intuitive (Figure~\ref{fig:appendixA-main}), often expressing confusion about which tactor was vibrating during navigation. 
This confusion likely diverted their focus from the navigation task itself, as they struggled to identify the source of vibrations. 
This observation aligns with prior studies demonstrating reduced sensitivity to simultaneous vibrations~\cite{shah-2019spatial, yeganeh2023discrimination, chen-2018effect, tajdari-2024sensitivity}. 

The difficulty in distinguishing the origin of concurrent vibrations, combined with their continuous nature, appears to have heightened participants' frustration.
This was further supported by post-study questionnaire responses indicating increased dissatisfaction (Figure~\ref{fig:appendixA-main}). 
To address this, alternative strategies such as the adjacent-pair method~\cite{satpute2019-fingersight}, which alternates vibrations between two axes at defined intervals, could reduce confusion and improve the overall user experience, including task efficiency.
Pairwise comparisons further revealed that participants traversed a significantly larger portion of the total distance within the critical area near the target when using the worst-axis first approach in the horizontal layout, averaging 82.4\%. 
This represents notably better performance compared to other layout and guidance approach combinations, likely attributed to the worst-axis first approach's more intuitive and precise guidance. 

Overall, these findings suggest that haptic navigation devices should avoid relying on continuous, simultaneous vibrations, even if they prove a theoretically direct path to the target. 
Such designs are particularly ill-suited for devices delivering feedback to limited areas of the body.
Instead, adopting approaches that emphasize clarity and reduce cognitive load, such as alternating or sequential feedback strategies, can significantly enhance user performance and satisfaction. 


\subsection{Guidance Metaphor}
The push metaphor led to significantly longer traveled distances and increased time to target, particularly in the vertical layout. 
In this configuration, the push metaphor resulted in significantly higher times to target compared to other combinations of layout and metaphor. 
Additionally, participants spent a significantly smaller proportion of their hand trajectory distance near the target in the vertical layout with the push metaphor, underscoring its inefficiency for navigation in such scenarios. 

These findings align with prior research comparing the push and pull metaphors in navigation tasks, which consistently showed that the pull metaphor supports more efficient navigation and is strongly preferred by users~\cite{jansen2004vibrotactile-push-pull, gunther2018-tactileglove}. 
The results of the post-study questionnaire further validate this, revealing that participants experienced significantly higher frustration levels with the push metaphor (Figure~\ref{fig:appendixA-main}). They also found the push metaphor less intuitive and more challenging for precise target navigation. 
Given these results, the pull metaphor is recommended for vibrotactile navigation devices, regardless of the scene layouts.

\subsection{Distance Cues}
Based on both objective and subjective metrics, we found no statistically significant effects of the intensity mode on any dependent variables. 
Similarly, prior research using a haptic wristband for peripersonal object navigation also reported no significant effect of distance cues on time to target, though that study manipulated vibration period instead of intensity~\cite{shih2018-dlwv2}. 

These findings suggest a potential challenge in participants' ability to perceive distance cues effectively through intensity-based feedback. 
Post-study responses further supported this notion, as only one participant explicitly mentioned that intensity increases  were helpful for navigation. 
A potential explanation lies in the limited range of intensity adjustments, especially in the linear mode, where subtle changes may have been difficult for participants to detect.

In the zone mode, where intensity increased within a 7-cm radius around the target, participants might not have had sufficient time to perceive the escalation before reaching the target.
This issue could stem from the proximity of the target attainment threshold (r = 3.5 cm), which limited the time window for recognizing intensity changes.
Expanding the radius of the zone mode could yield different outcomes, particularly in scenarios involving larger search areas. 

Moreover, exploring alternative methods of encoding distance cues--such as varying vibration patterns instead of intensity--may offer more effective solutions.
Adjusting other vibrotactile parameters, such as frequency or rhythm, could enhance user perception and improve the communication of distance information.








     

\subsection{Sex Effects}

Male participants outperformed female participants in this study, demonstrating significantly shorter task completion times, reduced hand trajectory distances, and a higher percentage of hand trajectory within the critical area near the target (Figure~\ref{fig:trajectories}). Female participants also reported significantly higher levels of frustration throughout the study regardless of scene layouts, guidance approaches, and guidance metaphors (Figure~\ref{fig:appendixA-main}). 
These differences may be influenced by a range of factors, including potential sex-based sensitivity differences for vibrotactile feedback. 
However, the evidence for such differences remains inconclusive, as prior studies on vibrotactile sensitivity across sexes have yielded conflicting results. 
Moreover, a previous study by the authors, which employed the same tactor positioning as the haptic glove used in this research, found no significant sex effects in vibrotactile perception~\cite{tajdari-2024sensitivity}. It is worth noting, however, that this earlier study did not involve a navigation task, limiting the applicability of its findings to the present context.

The observed performance difference between males and females could instead be attributed to physical or behavioral differences that emerge during continuous vibrotactile feedback in navigation tasks. 
For instance, variations in hand size, upper extremity strength, or movement strategies may have influenced participants’ ability to interact with the haptic system. 

In addition, some studies highlight differences between males and females in navigation strategies in different environments. For example, a study on indoor navigation using written instructions found that males performed better with Euclidean cues, such as cardinal directions and metric distances, while females excelled with landmark-based instructions that relied on prominent features~\cite{saucier-2002sex}. However, few studies have explored these differences using vibrotactile cues in peripersonal navigation, highlighting the need for future research.
Moreover, males may have played more video games, as questionnaires from U.S. students and undergraduates revealed that males reported playing twice as much as females weekly~\cite{greenberg-2010orientations}. Given that some studies suggest self-reported gaming time influences navigation performance in virtual environments~\cite{yavuz-2024video}, this could partly explain why males outperformed females in the current study.





    
\subsection{Glove Design}
Participants generally had a positive experience with the glove, appreciating its comfort and minimal restriction of movement. However, some improvements could enhance usability, particularly in tactor placement. Many participants struggled to differentiate between the tactors b,c, and d (see Figure~\ref{fig:overall-system}). A study supports this finding, showing that vibrations near the wrist are less distinguishable than those in other areas, regardless of the vibration pattern (simultaneous vs. successive)~\cite{luzhnica2017vibrotactile}. 
The tactor locations in this study were selected based on prior work by the authors~\cite{tajdari-2024sensitivity}, which demonstrated high accuracy in identifying vibrating tactors with different temporal patterns. Additionally, the arrangement resembles the layout of keyboard keys (up, down, left, right), helping participants intuitively associate each tactor with a specific direction.
The size of the wrist compartment (which encapsulates the board) was also suggested to be reduced, and the vibration intensity adjusted to improve comfort. Despite these suggestions, participants provided highly positive feedback on the glove's potential to assist visually impaired users with peripersonal navigation or in low-visibility environments, as reflected in their responses to the post-study questions (Table~\ref{table:appendixB-open}).

\subsection{Limitations and Future Work}

This study has several limitations, with the primary challenge stemming from the motion capture equipment used (ZED2 depth camera). 
The body detection algorithm exhibited high sensitivity to variations in lighting and participant posture, resulting in errors and occasional missed detections during the study. 
To address these problems, participants were asked to wear darker clothing to enhance contrast between themselves and the background environment, partially mitigating detection challenges. 
However, this workaround was not a comprehensive solution.
Future studies could benefit from using more robust motion capture technologies, such as optical motion capture systems or inertial measurement units, depending on the resources and setup availability. 
These alternatives could provide more accurate and consistent tracking, thereby improving the reliability of navigation guidance.
Future studies should prioritize testing the haptic feedback design with the target user group, such as visually impaired individuals, rather than blindfolded sighted participants.
Visually impaired individuals often exhibit enhanced activity in brain regions associated with touch and tactile attention compared to sighted individuals~\cite{burton-2004cortical}, and they tend to use distinct navigation strategies~\cite{shafique-2024path}.
Testing with this population could provide more accurate insights into the effectiveness and practicality of the haptic feedback system.

Navigating to a virtual target versus a real object can differ in several ways, such as the sensory and spatial cues available during the task. Another limitation of this study was the lack of real objects as targets. In a previous study, similarly shaped cubes of different colors were used as targets in a comparable navigation task, with participants receiving vibrotactile feedback while blindfolded~\cite{wei2022object-haptic}. Future studies could adopt a similar approach to enhance the ecological validity of the device by simulating real-world scenarios, such as distinguishing between similar medication bottles on a shelf.



\section{Conclusion}
\label{conclusion}

In this study, we proposed and evaluated a haptic glove designed to aid peripersonal target navigation under limited vision conditions. 
The research examined the effects of scene layout, guidance approach, guidance metaphor, and tactor intensity mode on navigation performance. 
Key findings revealed that participants performed better in a horizontal layout compared to a vertical one, but the worst-axis first guidance approach and push metaphor outperformed in both layouts. 
Similarly, different intensity modes used for conveying distance cues did not significantly influence performance in any condition. 
Notably, sex differences were observed, with male participants generally outperforming females in task performance.
These findings provide actionable insights for designing haptic gloves optimized for peripersonal navigation tasks, particularly in scenarios involving limited vision. The proposed haptic glove has potential applications beyond the scope of this study, including navigation in augmented and virtual reality environments, and accessibility tools for individuals with motor impairments, such as Parkinson’s patients. Future research should explore the inclusion of visually impaired individuals as test participants and incorporate real-world target objects to further enhance the glove’s practicality and effectiveness.

\section*{Acknowledgments}
We acknowledge the contributions of JMU Wearable Computing Research Group, including Tyler Webster, Megan
Caulfield, Stephen Mitchell, and Justin Blevins.

\section*{Author Contributions}

M. T. was responsible for conceptualization, data curation, formal analysis, methodology, software, visualization, and writing the original draft. J. F. was responsible for the initial development of the hardware and software systems used in this study and for editing the final manuscript. S. L. was responsible for conceptualization, funding acquisition, investigation, methodology, resources, supervision, visualization, and writing (review and editing).














\section*{Funding}
This initiative was funded by 4-VA, a collaborative partnership for advancing the Commonwealth of Virginia.

\section*{Appendix A}
\label{appendix}

\setcounter{figure}{0} 
\counterwithin{table}{section}
\renewcommand{\thefigure}{A.1} 
This section provides a more extensive explanation of the mechanism used to generate vibrations toward the intended target position in each condition. The pseudo-code for this method can be found at the end of this section.

\subsection*{Calibration and Target Generation}
\begin{figure}[h]
    \centering
    \includegraphics[scale = 0.7]{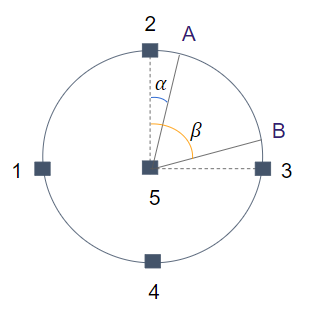}
    \caption{The captured points during the calibration procedure.}
    \label{fig:targets-appendix}
\end{figure}

Certain modifications should be implemented on the unprocessed footage captured by the camera in order to align the derived target positions with their actual locations on the circle. Initially, the circle may be captured as an ellipse rather than a perfect circle with equal distances from the top and sides, as a result of the camera angle. To accommodate for this potential variation, we recorded the top-most (2), right-most (3), and center (5) points of the circle during the calibration procedure, so that we could utilize them as a foundation for our target generation (Figure~\ref{fig:targets-appendix}). To capture each point, participants were instructed to hold their right palm at the specified location for 5 seconds. During this time, we recorded the position of the hand every 0.5 seconds. The final point position was calculated as the coordinate-wise median of these 10 points to mitigate any potential jitter in the skeleton model.

Potential camera setup issues may result in points 2 and 5 being misaligned vertically and points 3 and 5 being misaligned horizontally. Consequently, points A and B might be captured in place of points 2 and 3, respectively. Hence, another corrective action is done to map the range of 0-90 degrees linearly to the $\alpha$ to $\beta$ range (Figure~\ref{fig:targets-appendix}). In particular, for a desired target position at angle $\theta$, we generate the target at angle $\theta' = \frac {\theta (\beta - \alpha)}{90} + \alpha$.

After determining the adjusted angle, we find the adjusted target position around an ellipse. Let $d_{top}$ be the distance of the center point to the top-most point, and $d_{right}$ be the distance of the center point to the right-most point of the ellipse (distance of point 5 to point 2 and point 5 to point 3 in Figure~\ref{fig:targets-appendix}, respectively). We place the target at $(d_{right} \sin \theta', d_{top} \cos \theta')$.

\subsection*{Feedback Algorithm}

\paragraph{Motor positions}
We define unit vectors $\vec{v_a}, \vec{v_b}, \vec{v_c},$ and $\vec{v_d}$ in each of the four directions (up, down, right, and left) to assist in computing the vibration intensity of each motor. Note that when a user rotates their hand with a rotation angle of $\gamma$ with respect to the vertical line, the main axes also need to rotate to produce more accurate vibrations. This angle was calculated by creating a vector between the wrist and hand positions in the skeleton model, and then determining the angle between this vector and the vertical line. More precisely, we define the adjusted motor vectors as follows: (see Figure~\ref{fig:hand-rotation})
\begin{align*}
    \vec{v_a} &:= (\sin\gamma, \cos\gamma),&  \vec{v_b} &:= (-\sin\gamma, -\cos\gamma),\\
    \vec{v_c} &:= (\cos\gamma, -\sin\gamma),& \vec{v_d} &:= (-\cos\gamma, \sin\gamma).
\end{align*}

\renewcommand{\thefigure}{A.2} 
 \begin{figure}
            \centering
            \includegraphics[scale = 0.4]{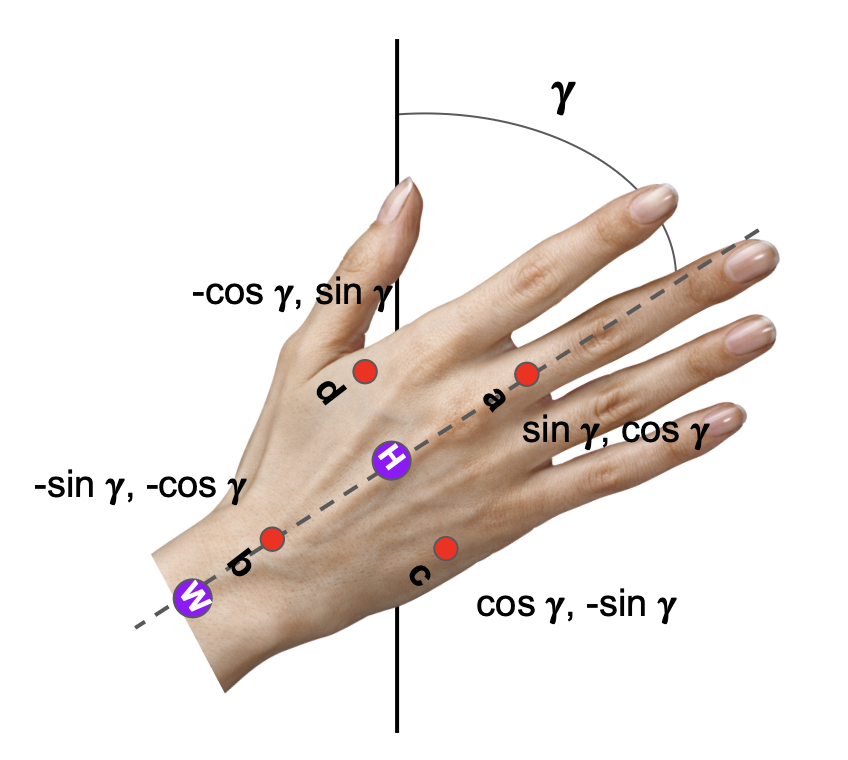}
            \caption{The adjusted vectors for each tactor based on the hand rotation angle ($\gamma$). H is the captured hand joint and W is the captured wrist joint in the skeleton model.}
            \label{fig:hand-rotation}
        \end{figure}

\paragraph{Determining the maximum vibration intensity} Let $d_c$ (resp. $d_h$) denote the distance of the center point (resp. user's hand) to the target. In the linear mode, if $d_h>d_c$ (i.e., the user's hand is farther than the center point to the target), then we set $I_{\max}=0.8$; otherwise, if $d_h\le d_c$, then we linearly increase $I_{\max}$ as $d_h$ decreases by setting $I_{\max}=1-\frac{0.2}{d_c}d_h$. In the zone mode, we define our zone as a circular area centered at the target that has a radius of $0.2d_c$, which in this experiment setup is equal to 7 cm. In this case, when $d_h>0.2d_c$ (i.e., the hand is outside of the zone), we set $I_{\max}=0.8$, and when $d_h\le 0.2d_c$ (i.e., the hand is inside the zone), we set $I_{\max}=1$. Note that the motors will start vibrating at an intensity of 0.59.

\paragraph{Determining the desired movement direction}
Next, we computed a unit vector $\vec{d}$ representing the direction in which we want the hand to move.
Let $\vec{v}$ be the normalized vector connecting the user's hand to the target position. In the two-tactor approach, we set $\vec{d}=\vec{v}$, indicating that the goal is to lead the hand directly toward the target. In the worst-axis approach, the direction \(\vec{d}\) is chosen to align with the axis that requires the most correction. Specifically, let
$m^* := \arg\min_{m \in \{a, b, c, d\}} \|\vec{v_m} - \vec{v}\|$,
where \(m^*\) identifies the motor with the greatest misalignment from \(\vec{v}\).
We set $\vec{d}=\vec{v}_{m^*}$.

\paragraph{Determining the motors vibration intensities} 
 Let $\vec{d}$ denote the direction computed by our method based on the selected mode. We determine how close is each motor to $\vec{d}$ by computing the distance of each motor \( m \in \{a, b, c, d\} \) as \( motor\_dist(m) := \|\vec{v_m} - \vec{d}\| \) (see Figure~\ref{fig:vector-positions}).
Note that these motor distances have values in the range $[0,2]$, where a motor distance of $0$ indicates that $\vec{d}$ coincides with the vector corresponding to the motor, a motor distance of $\sqrt{2}$ indicates that $\vec{d}$ is perpendicular to the vector corresponding to the motor, and a motor distance of $2$ indicates that $\vec{d}$ is exactly at the opposite direction of the vector corresponding to the motor. Using this interpretation, we then compute the amount of vibration of each motor as follows.  
\begin{itemize}
    \item In the pull mode, our glove needs to vibrate only the motors whose motor distance value is in the range $[0,\sqrt{2})$. Furthermore, for the motors with a motor distance in the range $[0,\sqrt{2})$, we increase the motor vibration linearly as the motor distance decreases, i.e., motors with $0$ motor distance are vibrated at $I_{\max}$ intensity and as the distance approaches $\sqrt{2}$, the vibration intensity decreases linearly until at distance $\sqrt{2}$, there will be no vibration. 
    \item In the push mode, our glove vibrates the motors that have a distance in the range $(\sqrt{2}, 2]$. Additionally, for the motors with motor distance in $(\sqrt{2}, 2]$, our glove increases the vibration intensity linearly as the motor distance increases, i.e., motors with $\sqrt{2}$ motor distance would not vibrate and as the distance approaches $2$, the vibration intensity increases linearly until at motor distance $2$, the vibration intensity would be $I_{\max}$. 
\end{itemize}

\renewcommand{\thefigure}{A.3} 
 \begin{figure}
            \centering
            \includegraphics[scale = 0.4]{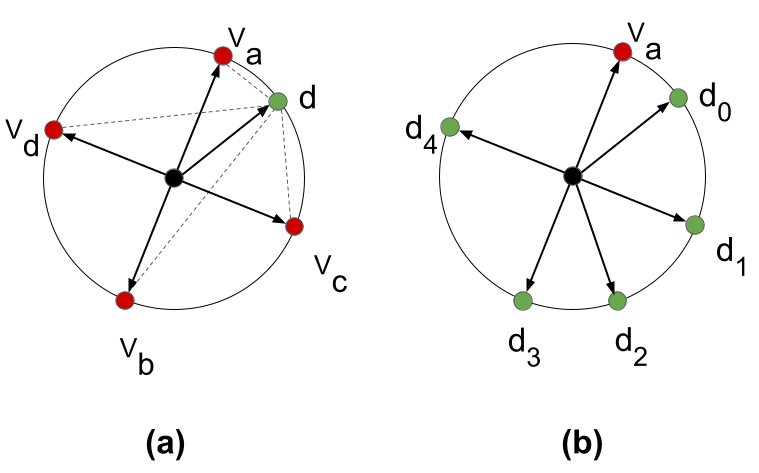}
            \caption{(a): The dashed lines demonstrate the motor distances respect to $\vec{d}$. (b): Motor distance of motor $a$ with respect to $\vec{d_0}$ is less than $\sqrt{2}$, with respect to $\vec{d_1}$ and $\vec{d_4}$ is $\sqrt{2}$, with respect to $\vec{d_2}$ is between $\sqrt{2}$ and 2, and with respect to $\vec{d_3}$ is 2.}
            \label{fig:vector-positions}
        \end{figure}
\newpage

   \begin{algorithm}[h]
    \scriptsize 
    \resizebox{\textwidth}{!}{ 
        \begin{minipage}{\textwidth} 
            \DontPrintSemicolon
            \caption{The vibration computation algorithm.}
            
            /* guidance approach: Worst-axis or two-tactor */
            
            /* intensity: Zone or linear */
            
            /* metaphor: Push or pull */
            
            /* center: The center of the circle. */
            
            /* target: Target position */
            
            /* hand: Hand position */
            
            /* $\gamma$: Hand rotation angle */
            
            \BlankLine
            \textbf{/* Motor positions */}\;
            $\vec{v_a} = ( \sin{\gamma}, \cos{\gamma}), \vec{v_b} = (-\sin{\gamma}, -\cos{\gamma}), \vec{v_c} = (\cos{\gamma}, -\sin{\gamma}), \vec{v_d} = (-\cos{\gamma}, \sin{\gamma})$\; 
            
            \BlankLine
            \textbf{/* Determining vibration intensities */}\;
            $d_h = \|target - hand\|$\;
            $d_c = \|target - center\|$\;
            \eIf{intensity = linear}
            {
            $I_{\max} = \max\{1 - \frac{0.2}{d_c}d_h, 0.8\}$\;
            }
            {
            \eIf{$d_h < 0.2d_c$}
            {
            $I_{\max} = 1$\;
            }
            {
            $I_{\max} = 0.8$\;
            }
            }
            
            \BlankLine
            \textbf{/* Determining desired movement direction */}\;
            $\vec{v} = \frac{target - hand}{\|target - hand\|}$\;
            \eIf{guidance approach = two-tactor}
            {
            $\vec{d} = \vec{v}$\Comment*[r]{Normalized vector of hand to target}
            }
            {

            $m^* := \arg\min_{m \in \{a, b, c, d\}} \|\vec{v_m} - \vec{v}\|$\;
            $\vec{d} = \vec{v}_{m^*}$\;
            }
            
            \BlankLine
            \textbf{/* Vibrating motors */}\;
            \For{each motor $m \in \{a, b, c, d\}$}
            {
            $\quad motor\_dist[m] = \|\vec{v_m} - \vec{d}\|$\;
            }
            
            \For{each motor $m \in \{a, b, c, d\}$}
            {
            \eIf{metaphor = pull}
            {
            $I[m] = \min\left\{0.59, I_{\max} - \frac{I_{\max} - 0.59}{\sqrt{2}}motor\_dist[m]\right\}$\;
            \Comment*[r]{Mapping $[0, \sqrt{2})$ to $[I_{\max}, 0.59)$}
            }
            {
            $I[m] = \min\left\{0.59, \frac{I_{\max} - 0.59}{2 - \sqrt{2}}motor\_dist[m] + \frac{1.18 - I_{\max}\sqrt{2}}{2 - \sqrt{2}}\right\}$\;
            \Comment*[r]{Mapping $(\sqrt{2}, 2]$ to $(0.59, I_{\max}]$}
            }
            }
            
        \end{minipage}
    }
\end{algorithm}

\newpage

\clearpage

\section*{Declaration of generative AI and AI-assisted technologies in the writing process}
During the preparation of this work the authors used ChatGPT in order to improve readability. After using this tool/service, the authors reviewed and edited the content as needed and take full responsibility for the content of the publication.
\newpage
\bibliographystyle{elsarticle-num} 
\bibliography{main}

\end{document}